# Evaluation of semi-monolithic scintillators with integrated RF shielding material for a higher integration of PET/MRI systems

**Emilia Laiyin Yin-Grossmann**[1], **Florian Mueller**[1], **Yannick Kuhl**[1], **Franziska Schrank**[1], **Stephan Naunheim**[1], **David Schug**[1,2] and **Volkmar Schulz**[1,2,3,4]

[1] Department of Physics of Molecular Imaging Systems, Institute for Experimental Molecular Imaging, RWTH Aachen University, Aachen, Germany
[2] Hyperion Hybrid Imaging Systems GmbH, Aachen, Germany
[3] Physics Institute III B, RWTH Aachen University, Aachen, Germany
[4] Fraunhofer Institute for Digital Medicine MEVIS, Aachen, Germany
E-mail: {laiyin.yin, volkmar.schulz}@pmi.rwth-aachen.de

August 22, 2022

**Abstract.**

Objective.
The integration of PET into MRI to form a hybrid system requires often compromises for both subsystems. For example, the integration might come at the cost of a reduced PET detector height or a reduced MRI examination volume diameter. Here, we propose a so-called shared-volume concept to use the volume required for both subsystems more efficiently, in which the MR transparency of the scintillator is exploited by integrating the RF shielding into the scintillator of the PET detector.

Approach.
Semi-monolithic scintillator prototypes (PVC slabs) of 7 mm and 12 mm height with integrated copper foil between every and every other slab, respectively, were investigated to evaluate the shielding effectiveness (*SE*). The *SE* was measured with probes on a test bench system. In addition, the PET detector performance was evaluated by determining the positioning using gradient tree boosting, energy and timing resolution using digital SiPM arrays (DPC3200, PDPC) in three LYSO scintillator configurations: ESR separation ($Slab_{ESR}$), ESR with copper foil in between ($Slab_{ESR+Cu}$), and purely copper foil separation ($Slab_{Cu}$).

Results.
The prototype with shielding between each slab and 12 mm height showed the highest *SE* with 31 dB (mean) in combination with a supporting frame. While $Slab_{Cu}$ was best for positioning, followed by $Slab_{ESR}$ and $Slab_{ESR+Cu}$, the $Slab_{ESR}$ had the best energy resolution (10.6 %), followed by $Slab_{ESR+Cu}$ (11.2 %) and $Slab_{Cu}$ (12.6 %). For the timing resolution, $Slab_{ESR}$ and $Slab_{Cu}$ achieved 279 ps and 288 ps, respectively. $Slab_{ESR+Cu}$ performed worst (293 ps).

Significance.
The scintillator-based RF shielding shows good RF shielding with similar PET performance, demonstrating the potential for more effective integration of PET detectors into MRI systems.

*Keywords*: RF shielding, integrated shielding, semi-monolithic, PET/MRI, machine learning, gradient tree boosting, DOI

# 1. Introduction

Merging positron emission tomography (PET) and magnetic resonance imaging (MRI) to a temporally and spatially matched imaging system requires an efficient utilization of the available space while maximizing each subsystem's sensitivity. The PET system must be optimized to run inside of a static magnetic field and needs to withstand the strong switching magnetic fields of the MRI [1, 2]. The selection of the used material is crucial as, e.g., magnetic components affect the $B_0$ homogeneity. Apart from the magnetization of the material, conductivity plays a key role during the selection process. The switching gradient fields induce eddy currents in conductive areas [3]. The resulting superimposing induced fields can affect the MRI's spatial encoding, and heating effects can occur inside of the PET modules. Material selection, size and structuring of the material can limit these effects, and materials with a low conductance and small area size are preferred for PET/MRI applications. While there are solutions for the optimization concerning gradient transparency [2], the radio frequency (RF) field poses a bigger challenge. The PET system needs to be shielded properly using a topology with high shielding effectiveness (*SE*) to reduce the impact of the strong and fast switching RF field on the PET electronics and vice versa.

Conventionally, this is done by encasing the whole PET module (electronics, photosensor and scintillator) in an RF shielding and light-tight housing (see figure 1(a)). Typical materials are, e.g., copper plates or foils [1, 4-7], carbon fiber [2, 8, 9] or laminated meshes [10] that have a good gradient compatibility (kHz range) and provide a high *SE* at the Larmor frequency (MHz range, e.g., 300 MHz for a 7 T MRI system). Shielding can be implemented module-wise [1, 4-7, 11, 12] by shielding every module individually, or system-wise [7, 9, 13] using, e.g., two tubes between which the modules are placed. Concerning the integration position, previous setups have shown different solutions: The PET system can be integrated inside the bore of an existing MRI Tx/Rx coil. Here, a high demand for efficient shielding exists as the PET modules are directly exposed to, and could potentially be damaged by, the RF magnetic field, or could also inflict damage to the MRI itself [11, 12, 14]. Another solution is the integration between the MRI coil's RF screen and the gradient system [15-17]. The RF screen of the resonator acts as an additional protective layer, and the PET system is less exposed. In this case, the MRI coil must be custom designed to provide sufficient space for the PET system by reducing the RF screen size. This process is highly individual as it depends on the used PET components and MRI type. However, a reduction of the RF screen diameter leads to a reduced coil sensitivity which lowers the SNR of the MRI system [18, 19]. Thus, the two subsystems compete for space when integrated as a single imaging system.

Reducing the RF shielded volume for the PET is, thus, favorable to increase the performance of the combined system while using the available space most efficiently. All PET module components are, therefore, built as compact as possible, but are limited by the trade-off between the subsystems' performances. One approach that improves this trade-off is to exclude the scintillator from the shielded volume: The exclusion of the scintillators from the RF shielding housings of the PET detector stack is possible as they do not interfere with the strong magnetic fields of the MRI and show a similar susceptibility as human tissue and almost no conductivity [20]. This exclusion reduces the housing volume and, depending on the placement of the PET ring, the occupied RF volume, which can be beneficial for the MRI sensitivity [18,19]. Additionally, the scintillator height can be increased which is of interest for the PET system's performance. However, the optical connection between scintillator and photosensor needs to be ensured while providing a sealed RF shielded volume.

Prior studies tested a shielding concept based on optically transparent, non-magnetic RF shielding materials. The shielding material was inserted between the photosensor and the scintillator (see figure 1(b)) [21,22]. The housing size was reduced by excluding the scintillator and connecting the integrated shielding material to the RF housing. Thus, the housing could be minimized to the height of the PET electronics and cooling structure. As the material would occlude the sensor area and could lead to an attenuation of optical photons, it needed to provide a high transparency to maintain a good PET performance.

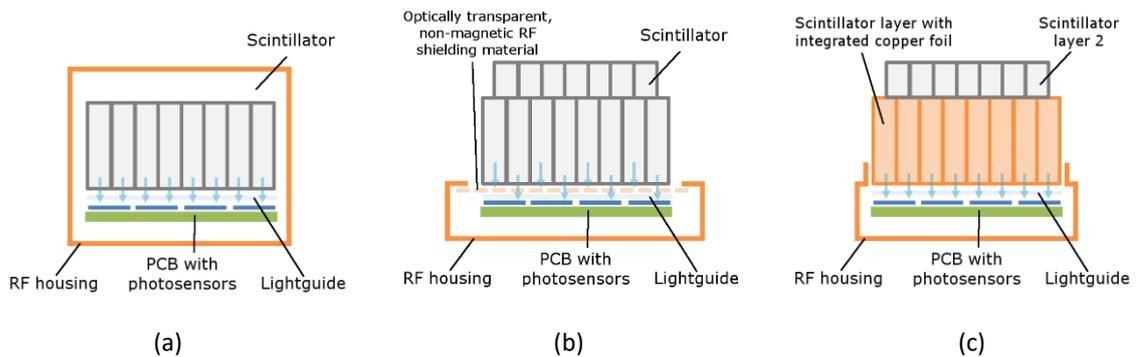

**Figure 1.** PET module with different RF shielding configurations: Each module consists of a scintillator array (grey), a light guide (light grey), and a photodetector (green PCB with blue sensors). The shielding is given in orange and the blue arrows indicate the path of the optical photons from the scintillator to the photosensor. (a) Conventional shielding that encases the PET module. The RF shielding is depicted in orange. The shielded volume is restricted and limits, e.g., the scintillator size. (b) Shared-volume approach with optically transparent, non-magnetic RF shielding material inserted between scintillator and photosensor. Some of the optical photons are attenuated by the integrated material (see blue arrows). A bigger or more complex scintillator geometry is feasible as more space is available towards the top of the detector. (c) Shared-volume approach with RF shielding material inserted in the segmented scintillator. The material is integrated between the scintillator segments and an outer wrapping ensures the connection among the layers. Analog to the prior concept, a bigger scintillator can be chosen. The wrapping is limited to one scintillator layer to connect to the RF housing of the PET detector.

Several materials were found as suitable candidates depending on the system's required *SE*. However, these materials are coated and attenuate photons in varying strengths. Therefore, they impede the PET detector performance. Furthermore, connecting the materials to the rest of the PET housing can be challenging as most of them are not rigid and would require an elaborate connection solution.

We propose an integrated RF shielding topology with RF shielding material integrated into the scintillator to address the challenges of the former shared-volume approach [21, 22]. We reduce the optical photon attenuation by integrating shielding material directly into the scintillator, or precisely, between the scintillator segments. This is done by wrapping scintillator segments with shielding material like, in our case, thin copper foil. The structure acts as a waveguide whose cut-off frequency is set above the MRI's Larmor frequency by adjusting the dimensions of the structure. Thus, the integrated shielding attenuates the penetrating or emitted electromagnetic fields. The integration is feasible for segmented scintillators (pixelated or semi-monolithic (slab) scintillators) as the segments are conventionally wrapped in a highly reflective material like, e.g., enhanced specular reflector (ESR) foil [1, 23] or a barium sulfate coating [24]. Especially for the ESR foil wrapping, an easy and clean integration of thin copper foil would be possible by wrapping shielding material around division of the scintillator, i.e., ESR-wrapped segments. The photon attenuation decreases as less sensitive sensor surface is occluded in comparison to conventional shielding material (e.g., coated glasses) located between scintillator and photosensor. Additionally, gaps between the read-out channels (e.g., for the digital silicon-photomultiplier dSiPM arrays PDPC DPC-3200-22-44 digital silicon-photomultiplier at the position of the bond wires) can be utilized to integrate the material above less sensitive sensor areas. The assembled detector with integrated shielding can then be pushed through the aperture in the PET RF housing. The connection between both components is realized through, e.g., conductive tape or gaskets (see figure 1(c)).

In this work, we will focus on semi-monolithic scintillators and compare the influence of copper foil integration on the PET detector performance using three scintillator array configurations. Semi-monolithic scintillators spread the light along one dimension over multiple read-out channels. This scintillator type provides depth-of-interaction (DOI) information with a good spatial resolution and high optical photon density [25]. It was shown that they have a good time, energy, and spatial resolution [26]. Recently, we presented a study with ESR-wrapped slabs using fan-beam irradiation and a gradient tree boosting (GTB)-based evaluation targeting clinical applications [27].

The semi-monolithic scintillators with integrated shielding material enable a PET RF shielding concept that could be suitable for applications with MRI volume coils, i.e., birdcage resonators: The typical current distributions along the birdcage rods lead to electromagnetic fields dominant radially around the rods, which leads to an RF eddy current that is mostly running along the direction of the birdcage rods. Thus, we expect a higher demand for shielding that supports the RF eddy current dominantly in axial direction.

However, the introduction of the copper foil needs to be evaluated to assess the achievable *SE*. The shielding material can change the reflectivity of the scintillator separation and, consequently, influence the PET detector performance. Especially the long-term stability is of interest as, e.g., copper foil oxidizes over time, which could increase the impact on the performance.

We evaluate the feasibility of this RF shielding approach using PVC prototypes and semi-monolithic lutetium-yttrium oxyorthosilicate (LYSO) slabs. The evaluation is conducted with respect to the achievable *SE*, and the influence of the introduced shielding material on the PET detector performance using dSiPM arrays and scintillator array configurations.

## 2. Materials

A summary of the used prototype and scintillator array configurations is given at the end of subsection 2.2.2 in table 1.

### 2.1 Shielding effectiveness setup
We used prototypes of the semi-monolithic detector configurations for the *SE* assessment. In the following subsections, the prototypes and the setup are described in detail.

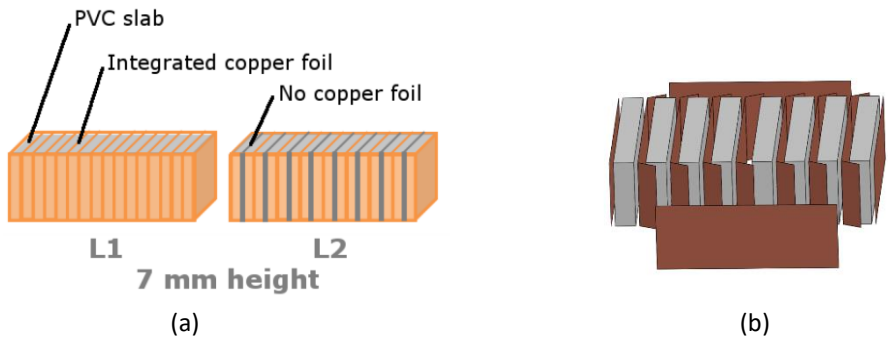

**Figure 2.** Prototypes for the *SE* evaluation. (a) 2 × 32 mm² PVC slabs with a height of 7 mm (low, L, depicted) or 12 mm (high, H) were used to build the prototypes. The copper foil was inserted between each (H1, L1) or every second slab (H2, L2). (b) Exploded view of a PVC prototype for *SE* evaluation: Example view of L1 (copper between every slab and 7 mm height). Copper foil (brown) was added between every slab and extruded a bit to fold against the slab. The outer wrapping is implied by the four outer copper sheets in the exploded view and connects all layers.

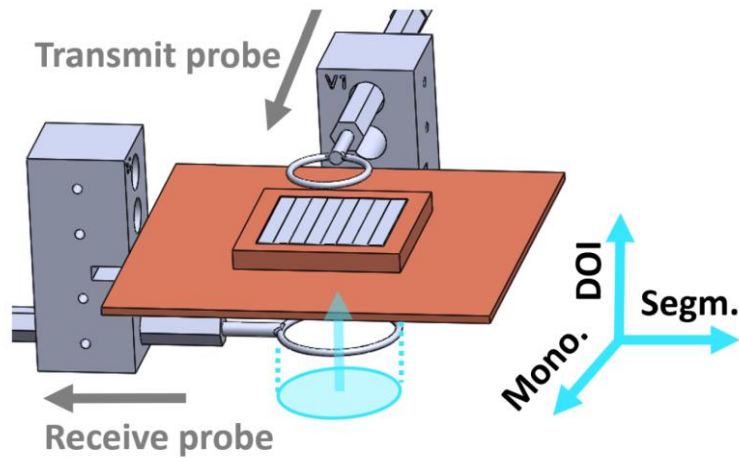

**Figure 3.** *SE* measurement setup: The prototypes were inserted in the frame (orange). Gaskets were inserted in a notch of the frame to ensure a conducting connection. The probe surface normal provided the receive probe orientation (blue) and was oriented in DOI direction for all measurements. The receive probe was stepped in a grid with 4 mm stepping width using a linear stage.

*2.1.1 Prototype configurations*

For the evaluation of the *SE*, prototypes of the integrated scintillators were assembled using 16 PVC slabs (2 × 32 mm²) with a height of 7 mm or 12 mm. Copper foil with 12.5 µm thickness was inserted between every, or every second slab, resulting in four combinations (height, copper foil position, see figure 2(a)):

- H1: 12 mm; every slab
- H2: 12 mm; every second slab
- L1: 7 mm; every slab
- L2: 7 mm; every second slab

The copper foils extruded to the lateral sides of the PVC slabs and were folded against the slabs (see figure 2(b)). An outer wrapping connected copper layers to ensure proper conductivity.

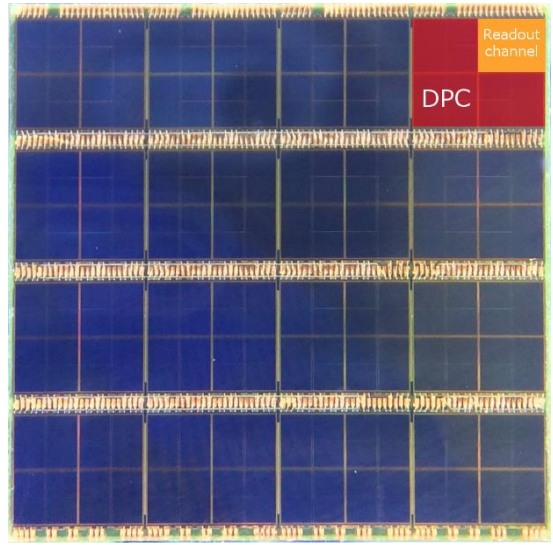

**Figure 4.** Photo of the PDPC DPC-3200-22-44 dSiPM sensor array: Each sensor array has 16 individual DPC 3200-22 sensors (DPCs, red). The DPCs have four readout channels each (orange). The area containing bond wires and connection padsis visible as golden regions between the DPCs and at the edges.

*2.1.2 Lab setup for SE evaluation*

The measurement setup consisted of an Agilent E5071C network analyzer (Santa Clara, California, USA) connected to two field probes (Langer EMV RF2, Bannewitz, Germany). A 30 dB pre-amplifier (Langer EMV PA303) was connected between the network analyzer and the receive probe. The receive probe was fixed to the frame and measurements were done for three orientations of the receive probe (monolithic, DOI and segmented direction) as shown in figure 3. The transmit probe was mounted on a linear stage (SMC5-USB-b9-2, Standa, Lithuania).

2.2 PET detector stacks

The PET detector stack comprised a photosensor and a semi-monolithic scintillator array. The components are introduced in detail in the following.

*2.2.1 Photosensor*

For the photosensor, dSiPM arrays (DPC-3200-22-44, Philips Digital Photon Counting, PDPC) with 16 individual and self-triggering DPC 3200-22 sensors, called DPCs, were used (see figure 4). Each DPC has four readout channels with a pitch of 4 mm, which leads to a total of 64 readout channels per sensor array. Every readout channel has 3200 single photon avalanche diodes (SPADs) that are connected to an individual logic circuit and can be divided into four sub-channels used for trigger generation. Bond wires connect the DPCs to the PCB and route signals to a central FPGA. The area containing these bond wires and the connection pads is visible as golden regions in figure 4 with a width of about 0.8 mm.

Four trigger schemes are available (trigger 1 - 4), which describe the average trigger threshold to start the acquisition scheme. They are set using Boolean operators (AND, OR) between the four sub-channels. The lower trigger schemes are more susceptible to triggers caused by dark count events, but the timing performance is better [28]. After triggering, an acquisition sequence containing validation, integration and recharging of the cells is started.

A detailed description of the trigger schemes, resulting average photon threshold and the acquisition sequence is given in [29].

Upon individually triggering and validating, the DPC sends a data packet (hit) towards an FPGA. Each hit contains the photon count value per readout channel and a timestamp. One gamma photon interaction could result in up to 16 hits, i.e., 16 timestamps and 64 optical photon count values.

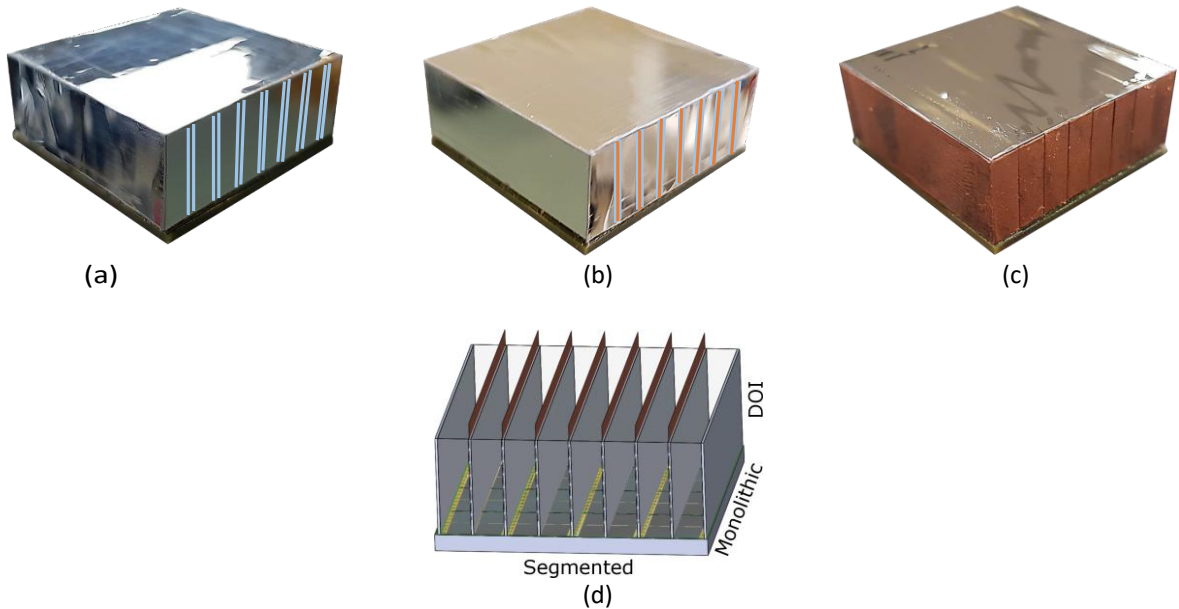

**Figure 5.** Scintillator array configurations. (a) Slab$_{ESR}$ : The slabs were separated with a double layer of ESR foil (indicated as blue double layers). The top and lateral sides were wrapped with ESR. (b) Slab$_{ESR+Cu}$ : 12.5 µm copper foil was inserted between the double layer of ESR foil (blue-orange-blue layers). The top and lateral sides were wrapped with ESR. (c) Slab$_{Cu}$ : Each slab was wrapped in 50 µm copper foil to match the pitch of the other detector configurations. The top was wrapped with ESR. (d) Detector orientation: We defined the segmented, DOI and monolithic direction as shown in the figure. The scintillator arrays' ESR and copper foils were aligned with the detector's bond wire gaps and each slab covered a row of DPC readout channels fully. The copper foils protrude from the array in the visualization to indicate the position. ((a)-(c) modified reprint with author's permission from [30]).

*2.2.2 Semi-monolithic scintillator*

Semi-monolithic scintillators combine the properties of monolithic scintillators and segmented arrays: they focus the optical photons on a smaller readout area and, thus, generate a higher photon density on the photosensors, and inherently provide DOI information unlike standard segmented scintillator arrays. The spatial calibration effort is reduced compared to monolithic scintillators as only the calibration in monolithic and DOI direction is needed. Apart from a good spatial resolution, they also have a good energy and timing resolution.

**Table 1.** Overview of the prototypes for the *SE* evaluation and the scintillator array configurations for the PET detector performance evaluation.

|  | Slab dimensions/material | | Separation |
|---|---|---|---|
| H1 | 2 mm x 32 mm x 12 mm | PVC | 12.5 µm copper foil, between every slab |
| H2 | 2 mm x 32 mm x 12 mm | PVC | 12.5 µm copper foil, between every second slab |
| L1 | 2 mm x 32 mm x 7 mm | PVC | 12.5 µm copper foil, between every slab |
| L2 | 2 mm x 32 mm x 7 mm | PVC | 12.5 µm copper foil, between every second slab |
| $Slab_{ESR}$ | 3.9 mm x 32 mm x 12 mm | LYSO | Double layer of 70 µm ESR foil |
| $Slab_{ESR+Cu}$ | 3.9 mm x 32 mm x 12 mm | LYSO | Double layer of 70 µm ESR foil with 12.5 µm copper foil in-between |
| $Slab_{Cu}$ | 3.9 mm x 32 mm x 12 mm | LYSO | 50 µm copper foil wrapped around every single slab with additional 50 µm copper foil sheet inserted in-between |

Eight 3.9 × 32 × 12 mm$^3$ LYSO slabs were used to assemble the scintillator array corresponding to the H2 configuration (copper foil between every second PVC slab), as the PVC slabs had half of the LYSO slab width. Every slab covered the area of one readout channel row (one-to-one coupling scheme), and the monolithic direction was matched along the bond wire gaps. The orientation can be seen in figure 5(d). Three scintillator arrays were tested as detector configurations:

- $Slab_{ESR}$: Every slab was separated with a double layer of ESR foil (70 µm per foil). The lateral sides were wrapped with ESR foil (see figure 5(a)).
- $Slab_{ESR+Cu}$: Every slab was separated with a double layer of ESR foil with 12.5 µm copper foil inserted in between. The lateral sides were wrapped with ESR foil (see figure 5(b)). This configuration has no electrical connection between the copper foil layers and is used to demonstrate the performance of a combination of ESR and copper foil for a better understanding.
- $Slab_{Cu}$: Every slab was wrapped using 50 µm copper foil. Another 50 µm sheet was added between the slabs to maintain the crystal pitch. The lateral sides were wrapped and did not require additional ESR foil (see figure 5(c)).

All configurations had an ESR top cover with 70 µm thickness. The scintillators were then mounted on the dSiPM sensors using an optical alignment tool to make each assembly as reproducible as possible. Both detector components were aligned to a global reference. The base of the scintillator array was coupled to the sensor array using optically clear adhesive foil and a silicone dielectric gel (Sylgard 527 silicone dielectric gel, Dow Corning, Midland, Michigan, USA).

The PET detector performance of $Slab_{Cu}$ and $Slab_{ESR+Cu}$ was compared to $Slab_{ESR}$ to evaluate influences of the introduced copper foil. These detectors were used as detectors under test in the following sections.

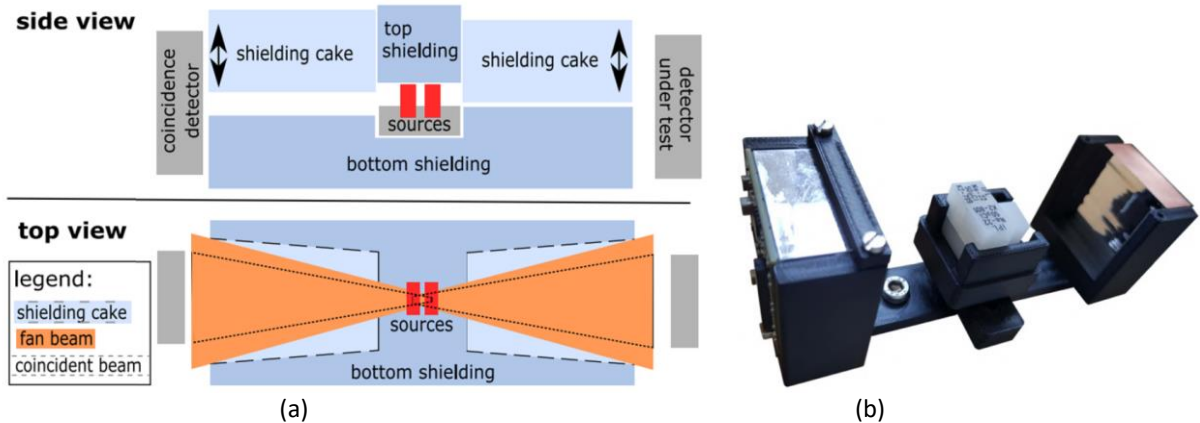

**Figure 6.** PET measurement setups. (a) Fan-beam-collimator with side and top view: The side view shows the coincidence detector and the detector under test at the sides of two shielding cakes. The latter define the beam width. In the top view, the restriction of the beam width is shown. Reprint from [31], CC3.0. (b) Flood measurement setup with sources and detectors: The two detectors face each other with five $^{22}$Na sources in-between. Reprint with author's permission from [30].

*2.2.3 Coincidence detector*

The coincidence detector had 8 × 8 LYSO scintillators with 12 mm height. The crystals were separated by ESR foil and coupled one-to-one to the dSiPM array. The top of the array was also covered with one layer of ESR foil. This detector was used as reference detector for all PET measurements.

2.3 PET measurement setup

We evaluated the PET detector performance regarding positioning, energy and timing resolution. For the positioning calibration of the monolithic and DOI direction, an external reference based on a fan-beam-collimator was employed. Energy and timing calibration relied on a flood calibration setup without collimator. The PDPC technology evaluation kit (TEK) was utilized as readout infrastructure. The setups were placed in a light-tight chamber.

*2.3.1 Measurements based on calibrations with fan-beam-collimator*

A fan-beam-collimator setup was used for the positioning calibration (see figure 6(a)): Two $^{22}$Na sources with 5 MBq activity each (Eckert & Ziegler, Berlin, Germany) were placed inside of the fan-beam-collimator setup. The sources were encased in lead shielding segments (top, bottom, side and "cakes"). The fan-beam was restricted to one side by the top and bottom shielding. The fan-beam width was then adjusted by the shielding cake.

Two detectors were set in coincidence in a distance of 240 mm. The coincidence detector remained in the same orientation and position for all measurements, and the shielding cake next to it was set to a fan-beam slit width of 7 mm. As this detector was used to determine the coincident beam, it was positioned closer to the collimator than the detector under test to reduce the impact of geometrically induced coincidence loss. Each semi-monolithic detector configuration was used as detector under test and was moved using a stepper motor (Limes 90, OWIS, Staufen im Breisgau, Germany). The shielding cake next to the detector under test was set to 0.4 mm, resulting in a fan-beam width of 0.533 ± 0.003 mm.

A more detailed description and evaluation of the setup itself can be found in [31, 32].

For positioning in the planar direction, the detectors were installed facing each other, i.e., the DOI direction of the detectors' orientation was parallel. The detector under test was stepped through the beam in monolithic direction and was measured from the crystal's top edge to the bottom edge. The top edge was found as described in [23, 33]. Afterwards, the edge of the detector was used as starting point and coincidence measurements are done in 0.75 mm steps.

For the positioning in DOI direction, the detector under test was rotated so that the fan-beam plane and the sensor surface area were parallel. The detector was moved in DOI direction for a lateral irradiation of the detector. The stepping width for the DOI measurements was set to 0.5 mm with the upper edge of the detector as starting point.

*2.3.2 Measurements based on flood irradiation*

A flood irradiation setup was used for the energy and timing calibrations (see figure 6(b)). Both detectors were placed in a 3D-printed setup with five $^{22}$Na NEMA cube sources (Eckert & Ziegler, Berlin, Germany) in-between. Four sources were stacked in a 2 × 2 matrix (about 0.3 MBq each), and the fifth source (1 MBq) was positioned centered next to them. The distance between the detectors and the sources' midplane was measured to the top surface of the detectors and corresponded to 40 mm for the coincidence detector, and 49 mm for the detector under test. The distance of the detector under test was higher to ensure coincidences all over the detector.

## 3. Methods

The evaluation was divided into two main segments: the evaluation of the *SE* using prototypes of the scintillator array configurations, and the evaluation of the PET performance using the three detector configurations for positioning, energy and timing resolution.

### 3.1 Evaluation of the shielding effectiveness

The prototype array segments were wrapped with copper foil to create waveguides. The aim was to attenuate the penetrating field by using a shielding structure that follows the principle of waveguides operating below cut-off frequency. The cut-off frequency is the frequency of the TE *i,j* mode, which is given for a waveguide with the dimensions *a > b* as [34]:

$$f_{c,ij} = \frac{1}{2\pi\sqrt{\mu\epsilon}} \sqrt{\left(\frac{i \times \pi}{a}\right)^2 + \left(\frac{j \times \pi}{b}\right)^2} \tag{1}$$

The cutoff frequency is given for the lowest TE mode (TE 1,0) as

$$f_{c,10} = \frac{1}{2a\sqrt{\mu\epsilon}} \tag{2}$$

In our case, with *a* = 0.032 m and PVC slabs with a typical $\epsilon_r$ in the range of 2.5 to 6 and $\mu_r$ = (1 + χ$_v$) = (1 − 10.71 × 10$^{−6}$) ≈ 1 [35, 36], we have a cut-off frequency range of 2.96 GHz to 1.91 GHz.

Below the cut-off frequency, the field is attenuated and the structure can be used as RF shielding material and the *SE* depends on the dimensions of the structure (*a* and height). We assess the influence of the field for several points of the grid to evaluate that local fluctuations that could potentially damage the sensor. The transmit probe was stepped in an isotropic grid with 4 mm step width over the frame using the linear stage. The *SE* was then calculated as

$$SE[db] = meas._{with\ material} - meas._{reference}. \qquad (3)$$

where the reference measurement had no shielding materials between the probes.

### 3.2 Pre-processing and calibration of PET data

Calibration data to evaluate the PET detector performance was acquired with the setups as described in section 2.3. All measurements were conducted in a light-tight climate chamber that was cooled to 0 °C for the collimator measurements (sensor temperature of 9 °C) and −5 °C for the flood measurements (sensor temperature of 3.5 °C).

In the following subsections, the methods for the position estimation in DOI and planar direction, and the energy and timing resolution estimation are described.

The data was recorded for all detector configurations with an inhibit fraction of 10 % to reduce the SPADs contributing most to the dark counts. The trigger scheme was set to 1 (triggering on the first detected optical photon) and the validation threshold was set to 4 (pattern 0x55:OR, 4-OR validation scheme in [29], corresponding to a validation limit of 17 ± 6.2 optical photons) with a validation length of 10 ns. All hits correlated to a single gamma photon interaction were clustered using a time window of 40 ns [37]. A coincidence time window of 10 ns was used. While no quality cuts were applied on the detector data during the processing steps, we report the positioning performance and timing resolution for an energy filter of 411 keV to 561 keV. A coincidence time window of 10 ns was used.

For both positioning estimations, the data was recorded at a bias voltage of 25.2 V with an overvoltage of 2.8 V and is split into training and test data using the number of events per irradiation position (training/test data):
- Planar: 10000 and 5000 events per irradiation position, respectively
- DOI: 5000 and 2000 events per irradiation position, respectively.

For the DOI model training and evaluation, an additional pre-processing step was done before splitting the data into training and test set. The lateral irradiation of the detector leads to a skewed distribution of the interaction positions according to the Lambert-Beer-law, which may introduce an additional bias to the data that misleads the machine learning technique during training. Thus, the distribution needed to be cut to assume an equal distribution which was then used for the training of the machine learning model.

Measurements were conducted for all three detector configurations (initial measurements). The stability of the copper wrapping surface over time was evaluated through measurements after two and four weeks to check the influence of oxidization and loss of reflectivity on the PET detector performance.

*3.2.1 Positioning estimation (fan-beam-collimator)*

The position was estimated as $y_i$ and $z_i$. In segmented direction, $x_i$ can be determined through the position of the readout channel with the highest photon count. The position estimations in monolithic direction ($y_i$) and DOI direction ($z_i$) were done using the supervised machine learning algorithm GTB and share the same methodology. A detailed description of the calibration routine and applied GTB algorithm can be found in [23, 31]. Briefly, GTB is utilized to build predictive regression models based on a set of labeled data (training data). The algorithm establishes an ensemble of sequential binary decisions (decision trees), which are evaluated as simple comparisons with two outcomes [38, 39]. The training process follows an additive manner: While the first decision tree is trained using the irradiation positions, every following tree corrects the already existing ensemble while focusing on the residuals (irradiation position – prediction of already trained ensemble) [39, 40]. Several hyperparameters influence both the positioning performance and computational requirements of trained GTB models. The most relevant are:

- Ensemble size: The number of established decision trees.
- Maximum depth: The maximum number of binary sequential decisions within one decision tree.
- Learning rate: The learning rate is a factor ranging from 0 to 1, applied to the residuals before training of the next decision tree.
- Input features: As shown in previous work, the positioning performance of GTB models benefits from adding additional, pre-calculated features motivated by the physical problem, e.g., first and second moment of the light distribution. In general, the utilized GTB implementation allows arbitrary input features and handles missing data caused by DPC having not triggered or validated intrinsically.

Trained GTB models can be implemented in several architectures, e.g., CPU, GPU, and FPGA [41, 42]. Recently, we have shown a high-performance CPU implementation for PET applications using the dedicated online processing platform for the Hyperion IID-platform [43].

Based on previous work [23, 31], we conducted a hyperparameter grid search to find the best performing GTB models without any restrictions on the computational complexity. We varied the number of trees up to 500, the maximum depth in range of {4, 6, 8, 10, 12}, and learning rates of {0.05, 0.1, 0.2, 0.4, 0.7} and utilized the data sets as described in section 3.2 for training and testing. During the training process, no energy filter was applied to the data, while for testing an energy window of [411, 561] keV was set. The GTB models were trained using the irradiation position as a label and the input features of each gamma photon interaction: the main DPC (DPC with the highest photon count value), the main readout channel (readout channel with the highest photon count value), and the photon sum (total sum of optical photons).

The estimated position $x_i$, $y_i$ and $z_i$ was then compared to the given true positions $Y$ and $Z$ of the fan-beam irradiation with the positioning errors $\Delta y = Y - y_i$ and $\Delta z = Z - z_i$.

The positioning performance was quantified based on the error distribution using:
- *SR*: The spatial resolution given by the *FWHM* of the error distribution (line

- spread function) [44],
- *MAE*: The mean absolute error, with $MAE(Y) = \frac{1}{N}\sum_i^N \|y_i - Y\|$
- *d*50 and *d*90: the distance from the irradiation position that covers 50 % and 90 % of the events,
- *RMSE*: Root mean squared error, with $RMSE(Y) = \sqrt{\frac{1}{N}\sum_i^N (y_i - Y)^2}$

The best performing model was selected using the *MAE* by using the *RMSE* metric as a loss function during training. The evaluation was done based on the calibration of the initial measurement to show the performance change expected within an assembled setup, and using new calibrations for each measurement to assess if the material changes can be corrected.

3.3 Energy calibration method and estimation (flood setup)

Flood calibration data was recorded with a bias voltage of about 25.0 V with an overvoltage of 2.8 V. The number of optical photons was corrected for saturation effects using a simple model for each readout channel [45]. For the energy calibration, a subset of clusters was selected while the energy resolution evaluation included all measured gamma interactions. To account for potential three-dimensional dependencies of the detector response, the scintillator volume is divided into subvolumes (voxels) for both energy calibration and energy calculation: The detector under test was divided into 8 × 8 × 4 voxels (i.e., four voxels along the DOI direction over each sensor readout channel). A histogram with the optical photon sum was created for each voxel to attain the photopeak value. To fill this histogram, only those clusters were selected fulfilling a neighborhood-criterion: Along with the hottest readout channel, the neighboring readout channels must be present for the current slab. Based on these clusters, the mean light pattern was calculated to interpolate missing readout channel values for the later energy calculation. The energy calibration was performed accordingly for the coincidence detector except by dividing the scintillator volume into 8 × 8 × 1 voxels following the utilized scheme of one-to-one-coupling. No neighborhood criterion was required here as the four readout channels of the main DPC represent a stable readout region.

The energy calculation utilized the sum of optical photons as defined before based on the neighborhood criterion accounting for missing photon values per cluster and applied the conversion factor found for the photopeak value. After the energy calibration, the energies were filled into a histogram. The energy resolution was obtained by fitting a Gaussian function.

3.4 Timing calibration method and estimation (flood setup)

The timing resolution is a key performance parameter in PET detector evaluation, since a sufficiently good coincidence timing resolution (*CTR*) results in an improved SNR of the reconstructed PET image. Usually, the *CTR* is defined as the FWHM of the time difference distribution regarding a coincidence setup. The time difference $\Delta t_i$ of coincidence *i* is determined for two facing detectors (in the following text called m and n) with $\Delta t_i = t_{m,i} - t_{n,i}$ where *t* is the reported timestamp of the respective detector. These time differences are used to estimate the *CTR* using a numerical method as the two detectors are not of the same type and result in a slightly skewed distribution. A parabola fit considering the three highest histogram data points is used to estimate the maximum, from which then $t_l$ and $t_r$ at half of the maximum are derived, respectively for the left and right side. The *CTR* is then given as the difference of $t_l$ and $t_r$.

The detector's reported timestamps are affected by so-called time skews, representing additional time offsets on the true timestamps, resulting in a degradation of the *CTR*. These time skews occur due to electrical signal runtime differences between the 16 DPCs but also due to optical runtime differences between the readout channels and crystal sub-volumes. The timing calibration utilizes the flood irradiation setup (see figure 7(b)) and aims to reduce the influence of time skews.

Mathematically, the two facing detectors, called a and b, are separated into abstract channels, and each coincidence is assigned to a channel pair $(c_i, c_j)$, with $c_i \in a$ and $c_j \in b$. Based on the relations between two channels, a matrix equation can be established which uses the mean time difference $\Delta t_{ij}$ between two channels $c_i$ and $c_j$, and information about the channel combination itself encoded in a product of matrix M and channel vector $\vec{c}$,

$$mean(t_i - t_j) = \overline{\Delta t_{ij}} = (\overrightarrow{\Delta t})_{ij} \equiv (\underline{M} \cdot \vec{c})_{ij} = c_i - c_j + t^{off}_{global}. \tag{4}$$

The parameter $t^{off}_{global}$ of equation 4 accounts for a possible off-centered radiation source position in case that all emission sources used for the calibration have the same distance to each of the detectors, and is assumed to be equal for all possible channel combinations, thus, representing a first-order approximation.

After the matrix equation has been formulated, the calibration corrections $\vec{c}_{min}$ can be found by minimizing a loss function $L(\vec{c}; \underline{M}, \overline{\Delta t})$ where, for this case, a simple minimization of the Euclidean 2-norm was chosen,

$$\vec{c}_{min} = \mathrm{argmin}_{\vec{c}} \left\| \underline{M} \cdot \vec{c} - \overrightarrow{\Delta t} \right\|^2. \tag{5}$$

The corrected timestamps are obtained by adding the calibration corrections onto the reported timestamps of the sensor tile.

The final calibration scheme used here utilized several iterations called calibration stages of the above-described mathematical procedure. Within each calibration stage, a different kind of "channelization" of the detectors was used, such that it was possible to target different aspects of the time skews. A more detailed description of the procedure can be found in [27].

## 4.    Results

### 4.1 Shielding effectiveness

The *SE* was evaluated for 100 MHz and 300 MHz. In the following, the focus was set on the results and plots at 300 MHz as this corresponds to the Larmor frequency of a 7 T MRI system. As the perpendicular field induces most effectively into the detector electronics, we investigate the *SE* for the DOI polarization only (probe surface normal corresponds to DOI orientation).

Figure 7 shows the influence of the copper frame with and without the prototype configurations compared to the S21 coefficient without frame. Overall, an average *SE* of more than 25 dB could be observed. A mean *SE*$_{Ref}$ of 25.2 dB was seen for the setup where only the copper frame was used for measurements (see 7(a)). A circular pattern was visible with higher *SE* towards the edges (32 dB to 48 dB) and lower *SE* in the middle of about 15 dB. The corners reached 23 dB to 29 dB. The insertion of the prototypes with copper integrated between each slab (H1 and L1) into the copper frame affected the *SE* attenuation and stripe patterns were visible, with a higher *SE* region towards the left and right edge (see figure 7(b) and 7(d)). Minor influences of the circular pattern were visible for L1 towards the edges. Here, H1 achieved on average higher *SE* values than L1 (31.2 dB and 30.3 dB for the mean *SE*$_{H1}$ and *SE*$_{L1}$, respectively, and 21.4 dB and 19.7 dB at the center). *SE* was improved by up to 20 dB towards the left and right side for H1, and up to 17 dB for L1. H2 showed a similar circular pattern as seen in the reference evaluation, figure 7(a), with a mean *SE*$_{H2}$ of 30.4 dB (see figure 7(c)). The edges improved in a range of 6 dB to 18 dB to *SE* values of 31 dB to 42 dB, and the center by around 4 dB to 19.4 dB. L2 achieved the lowest mean *SE*$_{L2}$ among the four prototypes with 28.5 dB (see figure 7(e)). The circular pattern could partially be seen on the top, bottom, and right side. Towards the left, the *SE* values are lower than for the reference evaluation. The minimum *SE* increased by about 3 dB to 17.8 dB

A summary of the observed mean and minimum *SE* is given in table 2 with additional values for *SE* at 100 MHz. At 100 MHz, the values and *SE* visualizations are similar for all detector configurations.

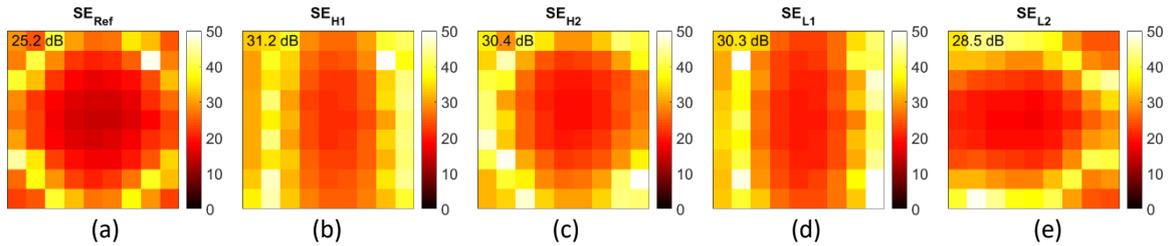

(a)     (b)     (c)     (d)     (e)

**Figure 7.** Visualization of *SE* over the stepping grid. (a) Reference: Measurement using the frame without inserted prototype. An average *SE*$_{Ref}$ of 25.2 dB could be achieved. (b) H1 prototype: The H1 prototype was inserted into the frame for this measurement. The average *SE*$_{H1}$ reached 31.2 dB (c) H2 prototype: Analog to H1 with an average *SE*$_{H2}$ of 30.4 dB. (d) L1 prototype: Analog to H1 with an average *SE*$_{L1}$ of 30.3 dB. (e) L2 prototype: Analog to H1 with an average *SE*$_{L2}$ of 28.5 dB.

**Table 2.** Comparison of the mean and minimum *SE* at 300 MHz and 100 MHz

|  |  | mean | | min | |
|---|---|---|---|---|---|
|  |  | 300 MHz | 100 MHz | 300 MHz | 100 MHz |
| $SE_{Ref}$ | [dB] | 25.2 | 25.1 | 14.6 | 14.4 |
| $SE_{H1}$ | [dB] | 31.2 | 32.3 | 21.4 | 21.8 |
| $SE_{H2}$ | [dB] | 30.4 | 31.8 | 19.4 | 20.3 |
| $SE_{L1}$ | [dB] | 30.3 | 30.2 | 19.7 | 19.9 |
| $SE_{L2}$ | [dB] | 28.5 | 29.2 | 17.8 | 18.4 |

## 4.2 PET Performance

A summary of the acquired results for all detector configurations is given in table 3.

### 4.2.1 Positioning performance

The best global *MAE* was achieved for a maximum depth of 12, a learning rate of 0.1 and 500 trees for both planar and DOI positioning. The *MAE* distributions are shown in figure 8. All planar *MAE* distributions (see figure 8(a)) have a constant area in the center with rising edges towards the detector edges. Here, $Slab_{Cu}$ has the lowest MAE overall. $Slab_{ESR+Cu}$ has a similar trend as $Slab_{ESR}$, but shows a higher *MAE* in the center. The results are given in table 2 for *SR*, *MAE*, *d50* and *d90* in millimeters.

For DOI positioning, *MAE* is again best for $Slab_{Cu}$, whereas $Slab_{ESR+Cu}$ performed worst (see figure 8(b)).

**Table 3.** Results for planar and DOI positioning (*SR*, *MAE*, *d50*, *d90*), energy resolution (*dE/E*) and timing resolution (*CTR*).

|   |   | $Slab_{ESR}$ | $Slab_{ESR+Cu}$ | $Slab_{Cu}$ |
|---|---|---|---|---|
| $SR_{Planar}$ | [mm] | 1.49 | 1.69 | 1.45 |
| $MAE_{Planar}$ | [mm] | 1.10 | 1.15 | 1.03 |
| $d50_{Planar}$ | [mm] | 0.62 | 0.64 | 0.60 |
| $d90_{Planar}$ | [mm] | 2.33 | 2.42 | 2.11 |
| $SR_{DOI}$ | [mm] | 3.54 | 3.71 | 3.00 |
| $MAE_{DOI}$ | [mm] | 1.60 | 1.68 | 1.47 |
| $d50_{DOI}$ | [mm] | 1.19 | 1.28 | 1.08 |
| $d90_{DOI}$ | [mm] | 3.53 | 3.72 | 3.27 |
| $dE/E$ | [%] | 10.6 | 3.72 | 12.6 |
| $CTR$ | [ps] | 279 | 293 | 288 |

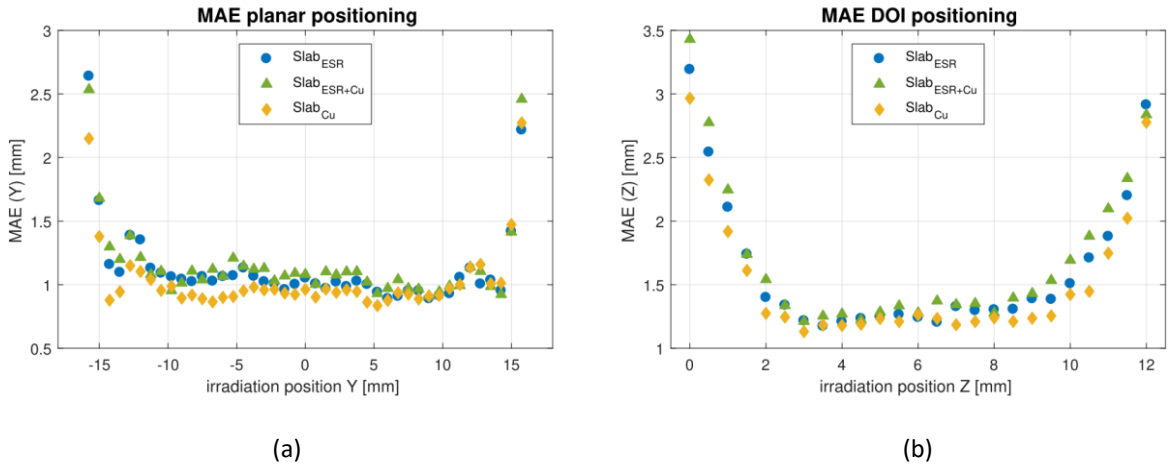

(a)      (b)

**Figure 8.** Mean absolute error (*MAE*) for the three detector configurations. (a) Planar positioning. The edges of the detector are at $Y = \pm15.75$ mm. (b) DOI positioning. $Z = 0$ mm corresponds to the detector surface and $Z = 12$ mm is the sensor plane.

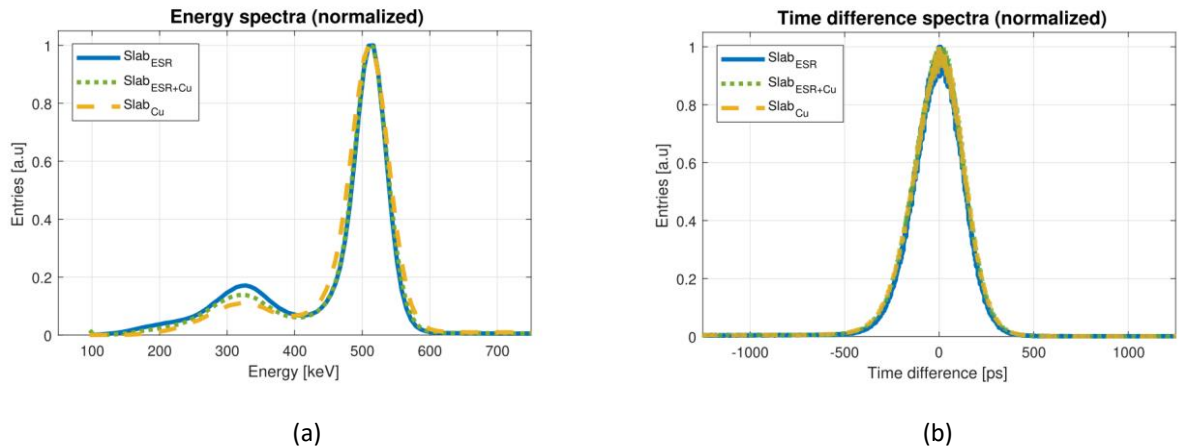

**Figure 9.** Normalized energy and time difference spectra for the three detector configurations. (a) Normalized energy spectrum. (b) Normalized time difference spectrum.

*4.2.2 Energy and timing resolution*

Slab$_{ESR}$ performed best with an *dE/E* of about 10.6 %, followed by Slab$_{ESR+Cu}$ and Slab$_{Cu}$ with 11.2 % and 12.6 %, respectively. The energy distribution is given in figure 9(a).

The lowest *CTR* was seen for Slab$_{ESR}$ with 279 ps. Slab$_{Cu}$ performed worse with a degradation from 279 ps to 288 ps. Slab$_{ESR+Cu}$ had the worst *CTR* with about 293 ps. The corresponding time difference spectrum is given in figure 9(b).

*4.2.3 Stability over time*

The detector configuration with only copper, Slab$_{Cu}$, was re-evaluated after two and four weeks to see how oxidization of the copper may affect the PET performance. The detector was stored in a climate chamber at temperatures between −5 °C and 15 °C during this time. The achieved performance parameters are shown in figure 10 (denoted by "long-term") and given in table 4. For initial calibration and planar positioning, the performance improved while for DOI a decrease was observed (figure 10(a) and (b)). The energy resolution decreased over the time (from 12.6 % to 13.1 %) but did not deteriorate much between the second and the fourth week (compare plots in figure 10(c)). The *CTR* degraded from 288 ps to 298 ps in the second week. At four weeks, the *CTR* degraded again to 293 ps. The corresponding time difference spectrum is given in figure 10(d).

The stability over time results shown in figure 10(a) to (d) were done using new calibrated position estimations for each data set. Using the calibration of the initial Slab$_{Cu}$ measurement showed an effect on the positioning performance (figure 10(e) for planar positioning and figure 10(f) for DOI positioning). Especially towards scintillator edges, a shift can be seen for both measurements which was are not prominent for the re-calibrated measurements in figure 10(a).

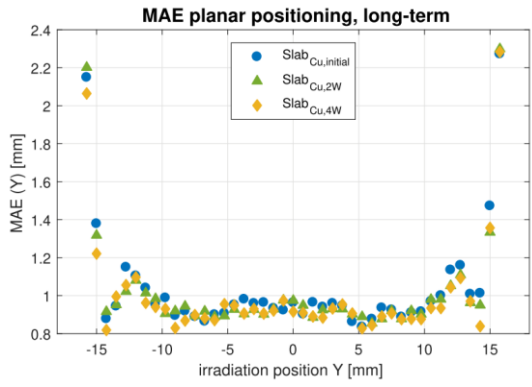

(a)

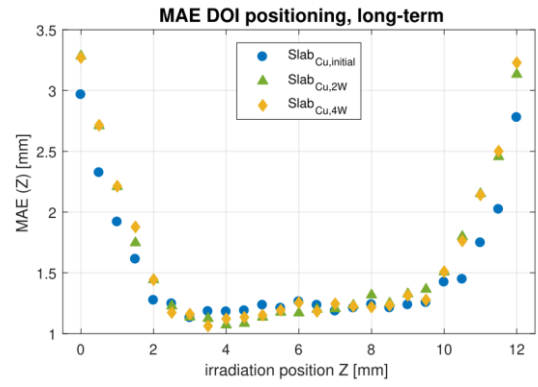

(b)

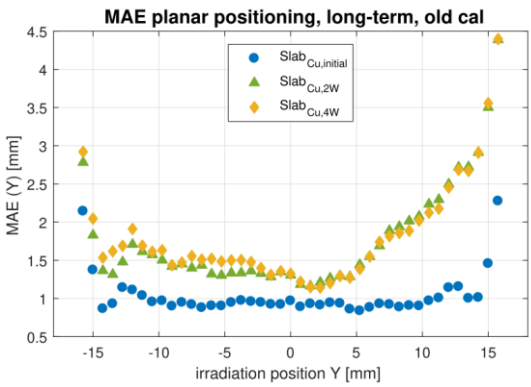

(c)

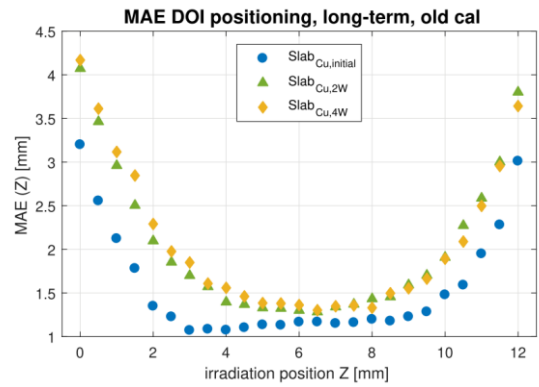

(d)

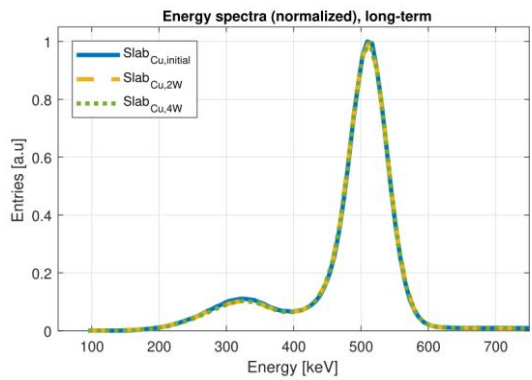

(e)

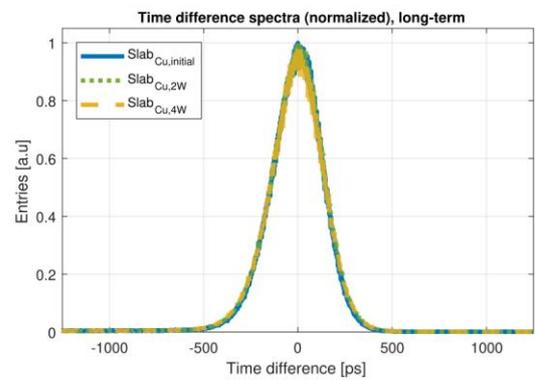

(f)

**Figure 10.** Detector performance comparison for $Slab_{Cu}$ for the initial measurement, after two weeks and after four weeks. (a) *MAE* for planar positioning. (b) *MAE* for DOI positioning. (c) Normalized energy spectrum. (d) Normalized time difference spectrum. (e) Stability of planar *MAE* when using the model of the initial $Slab_{Cu}$ measurement for all evaluations. (f) Stability of DOI *MAE* when using the model of the initial $Slab_{Cu}$ measurement for all evaluations.

**Table 4.** Results for the Slab$_{Cu}$ detector configuration. The results are shown for the initial measurement, after two and after four weeks, for planar and DOI positioning (*SR*, *MAE*, *d50*, *d90*), energy resolution (*dE/E*) and timing resolution (*CTR*).

|  |  | Slab$_{Cu}$ | Slab$_{Cu,2weeks}$ | Slab$_{Cu,4weeks}$ |
|---|---|---|---|---|
| *SR*$_{Planar}$ | [mm] | 1.64 | 1.55 | 1.56 |
| *MAE*$_{Planar}$ | [mm] | 1.03 | 1.01 | 0.98 |
| *d50*$_{Planar}$ | [mm] | 0.60 | 0.57 | 0.57 |
| *d90*$_{Planar}$ | [mm] | 2.11 | 2.07 | 2.00 |
| *SR*$_{DOI}$ | [mm] | 3.00 | 3.29 | 3.31 |
| *MAE*$_{DOI}$ | [mm] | 1.47 | 1.55 | 1.55 |
| *d50*$_{DOI}$ | [mm] | 1.08 | 1.15 | 1.16 |
| *d90*$_{DOI}$ | [mm] | 3.27 | 3.42 | 3.41 |
| *dE/E* | [%] | 12.6 | 13.1 | 13.0 |
| *CTR* | [ps] | 288 | 298 | 293 |

## 5. Discussion

### 5.1 Shielding effectiveness

Comparing the *SE* between 100 MHz and 300 MHz showed only minor differences. Hence, the discussion will focus on the 300 MHz.

The copper frame with the aperture for the prototypes attenuated the field by itself which was observable in the reference measurement (*SE*$_{Ref}$). A circular pattern was visible which re-occurred in the measurements using the H2 and L2 prototypes. The aperture itself has an impact on the *SE* distribution and should be kept small to increase the *SE* seen in the center.

The overall *SE* increased with the height of the copper wrapping and the insertion frequency (copper inserted between each slab vs. between every second slab) which aligns with our expectations concerning RF shielding with waveguides operating below cut-off frequency. The evaluation of the prototypes showed that the best *SE* performance at 300 MHz was achieved for H1 with a mean of 31 dB (prototype inserted in the copper frame). The overall *SE* increased slightly and was dominant to the sides. Both H2 and L1 achieved a mean *SE* of around 30 dB. However, the patterns seen in the *SE* visualization are different (H2 with circular pattern, L1 with stripe pattern). We suspect that the height of the shielding material has an effect on the resulting shielding area and *SE* is stronger along the edges of the monolithic orientation (compare figure 5(d) for orientation naming). This effect could be used as an advantage for slim apertures in which the SiPM arrays are not arranged in a square, but rather rectangular.

L2 showed a shift in the circular pattern. The achieved mean *SE* was lowest with around 28 dB. It is not clear whether the shift stems from the prototype geometry and the resulting field distribution, or from an assembly error.

In general, we have seen an increase in *SE* using the prototypes when measuring in DOI direction. This increase is rather low in comparison to the alternative shielding materials found in previous works for shared-volume concepts [22]. For example, the highest *SE* observed for the shielding approach using optically transparent, non-magnetic shielding materials was at 25 dB for a coated foil (HS9400, without a frame) [22] while the prototype H1 reached 31 dB at 300 MHz in combination with the frame. We have seen that the frame itself featured regions with high *SE* towards the edges with a lower *SE* region in the middle. A combination of an RF housing with a slim aperture and a scintillator with integrated shielding material might be a suitable solution for RF housings if the rest of the RF housing is made from materials with high *SE* properties, e.g., copper plates or foil. This would be suitable for setups requiring shielding in the range of 20 dB or slightly more for the given geometry (scintillator slab thickness of 2 mm, for a 32 × 32 mm$^2$ sensor area). In comparison, the PET detector configurations used for the PET detector performance evaluation use configuration H2. Since we are only closing the aperture created by the exclusion of the scintillator, high *SE* shielding material can be used for the rest of the housing (e.g., copper plates). By using even smaller scintillator divisions, an improvement of *SE* may be possible as the long side of the waveguide is the dominant factor for the shielding property. This can be done by, e.g., using two or more slabs per readout channel row. This would decrease the area in the center where a drop in *SE* was seen and a higher mean *SE* could be achievable.

As addressed before, the distribution of the field in an MRI's volume resonator is dominant around the resonator rods as the RF pulse needs to be perpendicular (xy-plane) to the orientation of the main magnetic field B$_0$ (z-axis). This indicates that a lower *SE* could be sufficient for fields along the z-axis and more shielding is required radially around the rods. The *SE* investigation in this work was simplified to assess the feasibility of copper integration as a PET RF shielding topology, and did not focus on how the orientation of the scintillator in the setup could be optimized. However, we estimate that the semi-monolithic scintillator with integrated RF shielding could have an orientation in which higher SE is observed, and we could utilize this by aligning the long edge of the waveguide structure with the z-axis of the MRI. A corresponding evaluation could be done with a prototype PET module featuring integrated RF shielding by performing MRI compatibility tests. The detector could then be rotated within the housing to see if one orientation improves the shielding.

The assembly of a module is easier with scintillators with integrated RF shielding and the presented shared-volume RF shielding approach in comparison to an integration of material between scintillator and photodetector. The coupling agent does not need to be spread through, e.g., a mesh [22], which could promote air inclusions, and couples the smooth surfaces of scintillator and sensors without occluding material in-between. By wrapping the scintillator segments with copper foil, a clean connection between RF housing and scintillator can be realized by using gaskets or copper tape. Disassembling and exchanging parts could benefit from this feature as well.

Thus, using shielding on the scintillator level to complete the RF shielding housing for the PET modules could be a suitable candidate for PET/MRI applications, given that a good connection to the PET module housing can be generated.

## 5.2 PET performance

The positioning results showed the best performance for $Slab_{Cu}$, followed by $Slab_{ESR}$ and $Slab_{ESR+Cu}$ for all evaluated performance parameters. The absorption of optical photons is highest for $Slab_{Cu}$ which leads to a better positioning performance. For $Slab_{ESR+Cu}$, the integration of copper between the ESR sheets could result in more optical photon scatter and, therefore, a worse performance is seen which is most prominent in the DOI position evaluation.

$Slab_{ESR}$ showed the best energy resolution with 10.6 %, followed by $Slab_{ESR+Cu}$ with about 11.2 % and $Slab_{Cu}$ with about 12.6 %. The decrease of the energy resolution stems from the increased optical photon absorption which increases with the thickness of the material.

The best timing performance was seen for $Slab_{ESR}$ with 279 ps, followed by $Slab_{Cu}$ with 288 ps. $Slab_{ESR+Cu}$ showed the worst *CTR* with about 293 ps.

Overall, $Slab_{Cu}$ showed a good PET performance concerning positioning and timing resolution, and the deterioration in both was minor in comparison to the reference detector configuration. In combination with the results from the *SE* evaluation, it could be a great candidate for PET/MRI applications with a focus on developing systems with spatial constraints. The biggest concern for the application of this shielding approach is the stability of the integrated material. As we use polished copper foil, we expect a performance decrease over time as the material oxidizes. The performance change seen for $Slab_{Cu}$ after two and four weeks is minor when using a new calibration per measurement. However, when using the initial calibration for the positioning estimation, we could see a degradation towards of the edges of the detector for both two weeks and four weeks. As this shift was seen primarily on one side for the planar *MAE*, we suspect that the effect stems from an irregular wrapping process during assembly instead of an overall aging process due to oxidization. Also, the trend is similar for both time points in comparison to the initial measurements. For the DOI MAE, a more uniform degradation is visible which could stem from the oxidation of the copper surface. Therefore, the detector needs to be observed over a longer time to make a definite statement concerning stability over time and aging effects. An evaluation to a later timepoint would be preferable to make a statement after a few years, mimicking the aging of a typical use time of multiple years in a PET setup. However, the results indicate an insufficient stability as the positioning performance deteriorated for the evaluation using the initial calibration. This would be a problem as detectors in an assembled setup are commonly not recalibrated in a fan-beam-setup in-between measurements.

We could only test one detector concerning its long-term stability as each configuration had to be disassembled to build the next. We think that $Slab_{ESR+Cu}$ could be better suited for system applications requiring a stable performance as the ESR layer could counteract the loss of reflectivity due to oxidization. However, the current configuration of $Slab_{ESR+Cu}$ uses two layers of ESR between which the shielding material is inserted. The shielding material is completely encased and cannot be connected to surfaces on the outside. By first wrapping the scintillator with ESR and subsequently with copper foil, the array would still have a double layer of ESR separation stemming from the neighboring segments, but also a double layer of copper foil in-between. The copper foil can then be connected from the outside of the array and complete the PET module's RF housing. This assembly procedure is not only feasible for single scintillator crystals but can also be extended to bigger segments of pixelated scintillators.

By using bigger scintillator segments, we can introduce shielding properties to pixelated scintillator arrays and adjust the waveguide size easily. This could be of interest for dSiPM arrays, like the one used in this work, but also for analog SiPMs like, e.g., the Broadcom AFBR-S4N44P163 analog SiPM array where shielding material can be integrated above the small gaps between the sensors. Therefore, we think that this shielding approach has great potential and should be further investigated.

## 6. Conclusion

We have introduced PET detector configurations with RF shielding integrated between scintillator segments and assessed the RF shielding properties using PVC slabs, and the PET performance using LYSO scintillator slabs coupled to dSiPM sensor arrays. The shielding performance ranged between 28 dB and 31 dB and could be suitable for applications with MRI volume resonators. Out of the four tested configurations, H1 with shielding between every slab and 12 mm height performed best (31 dB). For the PET performance, three detector configurations were tested: one with only ESR foil, one with ESR and a thin layer of copper foil, and one with a thicker copper foil as separation material for the slab detectors. The introduction of copper foil into the detector was successful and a minor impact on the PET performance could be observed. The positioning performance was best for only copper ($Slab_{Cu}$) and degraded only slightly for the energy (from 10.6 % for $Slab_{ESR}$ to 12.6 %) and timing resolution (from 279 ps for $Slab_{ESR}$ to 288 ps). Therefore, we are optimistic that using scintillator-based shielding is the next step to solve the challenges met while developing PET/MRI systems.

Future tests concerning the long-term stability of the $Slab_{ESR+Cu}$ detector could give more insight which of the two configurations is superior in a full system setup. Also, varying the wrapped volume by, e.g., choosing smaller slabs could be of interest for setups that require a higher *SE*.

**Acknowledgment**
This project is funded by the German Research Association (DFG), project number 288267690.

# 7. References

[1] B. Weissler et al. "A Digital Preclinical PET/MRI Insert and Initial Results". In: IEEE Transactions on Medical Imaging 34.11 (Nov. 2015), pp. 2258–2270. doi : 10.1109/TMI.2015.2427993.

[2] N. Gross-Weege et al. "Characterization methods for comprehensive evaluations of shielding materials used in an MRI". In: Medical physics 45.4 (2018), pp. 1415–1424.

[3] D. Le Bihan et al. "Artifacts and pitfalls in diffusion MRI". In: Journal of Magnetic Resonance Imaging: An Official Journal of the International Society for Magnetic Resonance in Medicine 24.3 (2006), pp. 478–488.

[4] C. Catana et al. "Simultaneous acquisition of multislice PET and MR images: initial results with a MR-compatible PET scanner". In: Journal of Nuclear Medicine 47.12 (2006), pp. 1968–1976.

[5] M. S. Judenhofer et al. "Simultaneous PET-MRI: a new approach for functional and morphological imaging". In: Nature medicine 14.4 (2008), p. 459.

[6] H. S. Yoon et al. "Initial results of simultaneous PET/MRI experiments with an MRI-compatible silicon photomultiplier PET scanner". In: Journal of Nuclear Medicine 53.4 (2012), pp. 608–614.

[7] J. Thiessen et al. "MR-compatibility of a high-resolution small animal PET insert operating inside a 7 T MRI". In: Physics in Medicine & Biology 61.22 (2016), p. 7934.

[8] J. M. Keith et al. "Shielding effectiveness density theory for carbon fiber/nylon 6, 6 composites". In: Polymer composites 26.5 (2005), pp. 671–678.

[9] B. J. Peng et al. "New shielding configurations for a simultaneous PET/MRI scanner at 7T." In: Journal of magnetic resonance (San Diego, Calif. : 1997) 239 (Feb. 2014), pp. 50–56. doi : 10.1016/j.jmr.2013.10.027.

[10] B. J. Lee et al. "Low eddy current RF shielding enclosure designs for 3T MR applications". In: Magnetic resonance in medicine 79.3 (2018), pp. 1745–1752.

[11] B. J. Lee et al. "MR performance in the presence of a radio frequency-penetrable positron emission tomography (PET) insert for simultaneous PET/MRI". In: IEEE transactions on medical imaging 37.9 (2018), pp. 2060–2069.

[12] A. M. Grant et al. "Simultaneous PET/MR imaging with a radio frequency-penetrable PET insert". In: Medical physics 44.1 (2017), pp. 112–120.

[13] S. H. Maramraju et al. "Electromagnetic interactions in a shielded PET/MRI system for simultaneous PET/MR imaging in 9.4 T: evaluation and results". In: IEEE Transactions on Nuclear Science 59.5 (2012), pp. 1892–1899.

[14] S. J. Hong et al. "SiPM-PET with a short optical fiber bundle for simultaneous PET-MR imaging". In: Physics in Medicine & Biology 57.12 (2012), p. 3869.

[15] J. Wehner et al. "MR-compatibility assessment of the first preclinical PET-MRI insert equipped with digital silicon photomultipliers". In: Physics in Medicine & Biology 60.6 (2015), p. 2231.

[16] V. Schulz et al. PET/MR scanners for simultaneous PET and MR imaging. WO Patent App. PCT/IB2008/050,046. July 2008.

[17] W. Branderhorst et al. "Evaluation of the radiofrequency performance of a wide-bore 1.5 T positron emission tomography/magnetic resonance imaging body coil for radiotherapy planning". In: Physics and imaging in radiation oncology 17 (2021), pp. 13–19.

[18] C. M. Collins et al. "A Method for Accurate Calculation of $B_1$ Fields in Three Dimensions. Effects of Shield Geometry on Field Strength and Homogeneity in the Birdcage Coil". In: Journal of Magnetic Resonance 125.2 (1997), pp. 233–241.

[19] K. Ocegueda et al. "A simple method to calculate the signal-to-noise ratio of a circular-shaped coil for MRI". In: Concepts in Magnetic Resonance Part A 28A.6 (2006), pp. 422–429. doi :

[20] S. Yamamoto et al. "Scintillator selection for MR-compatible gamma detectors". In: IEEE Transactions on Nuclear Science 50.5 (Oct. 2003), pp. 1683–1685. doi : 10.1109/TNS.2003.817375.

[21] C. Parl et al. "A novel optically transparent RF shielding for fully integrated PET/MRI systems". In: Physics in Medicine & Biology 62.18 (Sept. 2017), pp. 7357–7378. doi : 10.1088/1361-6560/aa8384.

[22] L. Yin et al. "RF shielding materials for highly-integrated PET/MRI systems". In: Physics in Medicine & Biology 66.9 (2021), 09NT01.

[23] F. Mueller et al. "A novel DOI positioning algorithm for monolithic scintillator crystals in PET based on gradient tree boosting". In: IEEE Transactions on Radiation and Plasma Medical Sciences 3.4 (2018), pp. 465–474.

[24] X. Zhang et al. "Development and Evaluation of a Dual-Layer-Offset PET Detector Constructed with Different Reflectors". In: Crystals 12.1 (2022). doi : 10.3390/cryst12010093.

[25] X. Zhang et al. "Performance of long rectangular semi-monolithic scintillator PET detectors". In: Medical physics 46.4 (2019), pp. 1608–1619.

[26] N. Cucarella et al. "Timing evaluation of a PET detector block based on semi-monolithic LYSO crystals". In: Medical Physics 48.12 (2021), pp. 8010–8023.

[27] F. Mueller et al. A Semi-Monolithic Detector providing intrinsic DOI-encoding and sub-200 ps CRT TOF-Capabilities for Clinical PET Applications. 2022. doi :

10.48550/ARXIV.2203.02535.

[28] D. Schug et al. "Crystal delay and time walk correction methods for coincidence resolving time improvements of a digital-silicon-photomultiplier-based PET/MRI insert". In: IEEE Transactions on Radiation and Plasma Medical Sciences 1.2 (2017), pp. 178–190.

[29] V. Tabacchini et al. "Probabilities of triggering and validation in a digital silicon photomultiplier". en. In: Journal of Instrumentation 9.06 (2014), P06016. doi : 10.1088/1748-0221/9/06/P06016.

[30] Y. Kuhl. "PET Performance Evaluation of Semi-Monolithic Scintillators with Integrated RF Shielding for PET/MRI Systems". Bachelor's Thesis. Fakult¨at f¨ur Mathematik, Informatik und Naturwissenschaften der RWTH Aachen, Sept. 2019.

[31] F. Mueller et al. "Gradient tree boosting-based positioning method for monolithic scintillator crystals in positron emission tomography". In: IEEE Transactions on Radiation and Plasma Medical Sciences 2.5 (2018), pp. 411–421.

[32] R. Hetzel et al. "Characterization and Simulation of an Adaptable Fan-Beam Collimator for Fast Calibration of Radiation Detectors for PET". In: IEEE Transactions on Radiation and Plasma Medical Sciences 4.5 (2020), pp. 538–545.

[33] C. Ritzer et al. "Intercrystal scatter rejection for pixelated PET detectors". In: IEEE Transactions on Radiation and Plasma Medical Sciences 1.2 (2017), pp. 191–200.

[34] D. M. Pozar. Microwave engineering. Fourth edition. John wiley & sons, 2011.

[35] P. Dobrinski et al. Physik f¨ur Ingenieure. Lehrbuch Physik. Vieweg+Teubner Verlag, 2010.

[36] M. C. Wapler et al. "Magnetic properties of materials for MR engineering, micro-MR and beyond". In: Journal of magnetic resonance 242 (2014), pp. 233–242.

[37] D. Schug et al. "Data Processing for a High Resolution Preclinical PET Detector Based on Philips DPC Digital SiPMs". In: IEEE Transactions on Nuclear Science 62.3 (June 2015), pp. 669–678. doi : 10.1109/TNS.2015.2420578.

[38] J. H. Friedman. "Greedy function approximation: a gradient boosting machine". In: Annals of statistics (2001), pp. 1189–1232

[39] T. Chen et al. "XGBoost: A Scalable Tree Boosting System". In: Proceedings of the 22nd ACM SIGKDD International Conference on Knowledge Discovery and Data Mining. KDD '16. San Francisco, California, USA: Association for Computing Machinery, 2016, pp. 785–794. doi: 10.1145/2939672.2939785.

[40] A. Natekin et al. "Gradient boosting machines, a tutorial". In: Frontiers in Neurorobotics 7 (2013). doi : 10.3389/fnbot.2013.00021.

[41] B. Van Essen et al. "Accelerating a Random Forest Classifier: Multi-Core, GP-GPU, or

FPGA?" In: 2012 IEEE 20th International Symposium on Field-Programmable Custom Computing Machines. 2012, pp. 232–239. doi : 10.1109/FCCM.2012.47.

[42] R. Kulaga et al. "FPGA implementation of decision trees and tree ensembles for character recognition in Vivado HLS". In: Image Processing & Communications 19.2-3 (2014), p. 71.

[43] C. Wassermann et al. "High throughput software-based gradient tree boosting positioning for PET systems". In: Biomedical Physics & Engineering Express 7.5 (2021), p. 055023.

[44] National Electrical Manufacturers Association and others "Performance measurements of small animal positron emission tomographs". In: NEMA Standards Publication, NU4-2008 (2008), pp. 1–23.

[45] D. Schug et al. "Initial PET performance evaluation of a preclinical insert for PET/MRI with digital SiPM technology". In: Physics in Medicine & Biology 61.7 (2016), p. 2851.